\documentclass[12pt]{article}

\usepackage{amsmath}
\usepackage{mathtools}
\usepackage{amssymb}
\usepackage{amsfonts}
\usepackage{amsthm}
\usepackage{mathrsfs}
\usepackage[all]{xy}
\usepackage[colorlinks=true, citecolor=blue]{hyperref}
\usepackage{tensor}

\usepackage{authblk}
\usepackage{fullpage}

\newcounter{mnotecount}

\newcommand{\mnotex}[1]
{\protect{\stepcounter{mnotecount}}$^{\mbox{\footnotesize $\bullet$\themnotecount}}$ 
\marginpar{
\raggedright\tiny\em
$\!\!\!\!\!\!\,\bullet$\themnotecount: #1} }

\newcommand{\dd}{\mathrm{d}}
\newcommand{\lie}{\pounds}

\newcommand{\beq}{\begin{equation}}
\newcommand{\eeq}{\end{equation}}

\newcommand{\az}[1]{{ #1}}

\DeclareMathOperator{\diag}{diag}

\def\al{\alpha}
\def\be{\beta}
\def\ga{\gamma}
\def\de{\delta}
\def\ep{\epsilon}
\def\ze{\zeta}

\def\si{\sigma}

\begin{document}

\title{Area deficits and the Bel-Robinson tensor}
\author[1]{Ted Jacobson\thanks{jacobson@umd.edu} }
\author[2]{Jos\'{e} M. M. Senovilla\thanks{josemm.senovilla@ehu.es} }
\author[1]{Antony J. Speranza\thanks{asperanz@gmail.com}   }

\affil[1]{\small \it Maryland Center for Fundamental Physics, University of Maryland, \mbox{College Park, 
MD 20742, USA}}
\affil[2]{\small \it F\'isica Te\'orica, Universidad del Pa\'is Vasco, Apartado 644, 48080 Bilbao, Spain}

\date{\today}

\maketitle

{\abstract
The first law of causal diamonds relates the  area deficit of a small ball relative to flat space to 
the matter energy density it contains.  At second order in the Riemann normal coordinate
expansion, this energy density
should receive contributions from the gravitational field itself.  In this work, we study the second-order area deficit of the ball in the absence of matter and analyze its relation to possible notions of gravitational energy.
In the small ball limit, any proposed gravitational energy functional should evaluate to the 
Bel-Robinson energy density $W$ in vacuum spacetimes.  
A direct calculation of the area deficit reveals a result that is not simply proportional to $W$.
We discuss how the deviation from $W$
is related to ambiguities in defining the 
shape of the ball in curved space, and provide several proposals for fixing these shape 
ambiguities.
}

\flushbottom
\pagebreak

\tableofcontents

\section{Introduction}
The Einstein field equations describe how the curvature of  spacetime  is related to 
the stress-energy of matter fields.  One way to understand this is through the effect that the 
curvature has on the volume of small spatial balls, or on the area of their enclosing boundaries.  Curvature 
causes a spatial ball of a given
volume to have a smaller surface area than it would in flat spacetime, 
and the Einstein equations state 
that this area deficit is proportional to the energy density at the center of the ball---with respect to the observer at rest with the ball \cite{Pauli1981, Feynman2003, Jacobson2015}.
This relation between area and energy density follows from an Iyer-Wald first law
applied to the ball \cite{Wald1993, Iyer1994a, Jacobson2015, Bueno2017}, 
which casts the Einstein equations in a thermodynamic light.  
In \cite{Jacobson2015} it was further proposed that this thermodynamic identity be interpreted
as an equilibrium condition for the vacuum entanglement entropy, 
establishing a novel principle for quantum gravitational theories.  

A natural question arises: how does the energy of the gravitational field  enter this picture? 
In particular, even
in the absence of matter, one might expect area deficits arising from purely gravitational effects.  
At lowest order in a Riemann normal coordinate (RNC) 
expansion about the center of the ball, the area deficit
is governed by the Einstein tensor at the center of the ball, and hence vanishes for a 
vacuum solution of the field equations.
However, since the Einstein equations are nonlinear, the gravitational field is itself a source of 
curvature, 
and this ``purely gravitational'' curvature will affect the area as well. 
Given the above relation between energy density and  area deficit 
in the presence of matter, 
a natural guess is that this change in area is related, in one way or another, 
to the gravitational energy density within the ball.
For this conjecture to hold weight, the area deficit should satisfy a number of properties associated with quasilocal
gravitational energy.  
To begin with, the change in area should be negative definite, corresponding to a positive gravitational energy.  
Furthermore, the ball is constructed relative to a timelike unit vector $u^\alpha$ at its center that defines a particular Lorentz
frame in which the energy is measured.  The energy should transform under a change in this frame
as the timelike component of some tensor field.  This tensor should be quadratic in the curvature at the center 
of the ball to match first contribution to the area deficit in vacuum.  These requirements already restrict us to considering
four-index tensors $t_{\alpha\beta\gamma\delta}$ that are quadratic in the Weyl tensor $C^\al{}_{\be\lambda\mu}$.

An additional desideratum for the underlying tensor $t_{\alpha\beta\gamma\delta}$ is that the putative energy  $t_{\alpha\beta\gamma\delta}u^\alpha u^\beta u^\gamma u^\delta$ propagate causally, in the sense that it vanishes in the entire domain of dependence of any region in which it vanishes.  This causal propagation is known to follow from a bound on the divergence of $t_{\alpha\beta\gamma\delta}$, together with the dominant property, which states that $t_{\alpha\beta\gamma\delta}$ contracted on any four future pointing vectors is non-negative \cite{B-Senovilla1997,Bergqvist-Senovilla1999}. The dominant property also guarantees that the ``momentum density"  
vector $p_\alpha = - t_{\alpha\beta\gamma\delta} u^\beta u^\gamma u^\delta$ is future-pointing timelike or null. This momentum density points in the direction of propagation of the putative energy  $t_{\alpha\beta\gamma\delta}u^\alpha u^\beta u^\gamma u^\delta$. 
The unique tensor \cite{Senovilla2000} with the dominant property and quadratic in the Weyl tensor is the Bel-Robinson tensor, defined in arbitrary dimension by
\beq \label{eqn:Tarb}
T_{\al\be\mu\nu} = C_{\al\rho\mu\sigma} C_\be{}^\rho{}_\nu{}^\sigma + 
C_{\al\rho\nu\sigma} C_\be{}^\rho{}_\mu{}^\sigma- \frac{1}{2} g_{\al\be}  C_{\tau\rho\mu\sigma} C^{\tau\rho}{}_\nu{}^\sigma 
-\frac{1}{2} g_{\mu\nu} C_{\al\rho\tau\sigma} C_\be{}^{\rho\tau\sigma}
+\frac{1}{8} g_{\al\be} g_{\mu\nu} C_{\zeta\rho\tau\sigma}C^{\zeta\rho\tau\sigma}.
\eeq
Furthermore, the tensor (\ref{eqn:Tarb}) is divergence-free in Einstein spaces. 
The associated Bel-Robinson ``super-energy'' density $W = T_{\al\be\ga\de}u^\al u^\be u^\ga u^\de$ \cite{Senovilla2000}
arises often in the context of quasilocal gravitational energy \cite{Szabados2009}, particularly when considering
small spheres \cite{HoSc1982, Bergqvist1994}, and it vanishes
if and only if $C_{\alpha\beta\gamma\delta}$ does.  
Because of the above properties and its relation to gravitational energy, $W$ is a natural 
guess for the quantity governing the second-order area deficit in the RNC expansion.

In this paper, however, we find that the area deficit for geodesic balls of fixed radius does not turn out to be proportional to $W$.
On the other hand, although the construction of a geodesic ball of constant radius is 
well-defined in both the flat and curved spacetimes, it may turn out that these are the wrong balls to 
compare when computing the change in local gravitational energy. 
In fact, the question of how to define the area variation 
appears already at first order in the presence of matter, where it is necessary to change the 
radius of the ball in curved space to ensure that its spatial volume  is held fixed 
\cite{Jacobson2015}.  The situation is more complicated at second order in perturbations, 
because the area becomes sensitive to shape deformations of the ball involving anisotropic 
deformations of the geodesic radius.   We should therefore 
expect to find an area variation proportional
to $W$ only for a special class of small balls, and only when comparing to the appropriate
ball in flat space, which may differ from a sphere.

Section \ref{sec:deformations} describes the effect that generic shape deformations have on the 
area and volume of the ball.  Most of these deformations yield the same contributions in curved
space as they do in flat space, so it seems unlikely that they 
could arise in a natural prescription for fixing the ball shape.
However, two types of deformations can yield contributions quadratic in the 
curvature:
a spherically symmetric change in the ball radius, and a spin-two deformation determined
by a symmetric, traceless, spatial tensor $Y_{ij}$, aligned with the electric-electric part \cite{Senovilla2000} of the 
Weyl tensor, $E_{ij}$---see section \ref{sec:EH} for the electric-magnetic decomposition of the Weyl tensor in arbitrary dimension.  We further argue in section \ref{sec:fixing} that a particularly natural
choice for $Y_{ij}$ arises by requiring that the domain of dependence of the ball  
lie in the interior of two intersecting lightcones, with caustics only at the apexes.  
Having made this choice of shape deformation, it is still necessary to determine the change in the 
overall radius of the ball, as well as to determine the ball in flat space with which to compare.  
Unfortunately, we were unable to give a fully satisfactory prescription for fixing these quantities;
however, we discuss in section \ref{sec:volume} some choices that lead to an area deficit
proportional to $W$.  We leave it as an open problem whether this or some other prescription
can be given a natural, geometric justification.

Ideally, the prescription for determining which quantities to compare would arise from a canonical
variational identity, 
as occurs at first order \cite{Jacobson2015}.
Progress in this direction is made in section
\ref{sec:noether} using the Noether charge formalism, which identifies a Hamiltonian associated with evolution
within the ball.
As long as the evolution vector vanishes at the ball's boundary and generates the Cauchy horizon of the ball,
the Hamiltonian is equal to the area on-shell, and once again differs from the Bel-Robinson energy density $W$.  
Again, we are faced with the issue of finding the appropriate ball in flat space with which to compare
the area.  This is related to the problem of fixing the overall constant by which the Hamiltonian
can be shifted without changing the dynamics.  One way to do this is by requiring 
that the Hamiltonian vanishes in flat space; however, this necessitates a prescription for 
determining the appropriate reference ball in flat space.  A similar issue arises in 
other definitions of quasilocal energy, such as that of Wang and Yau \cite{Wang2009}, 
which takes the 
flat space reference surface be isometric to the surface in curved space.
Finally, we also note that the Noether charge 
definition of this Hamiltonian is ambiguous \cite{Jacobson1994b}, 
and can be modified by terms depending on the extrinsic curvature
of the surface.  We show that there are choices of the ambiguity terms that cause the Hamiltonian to be proportional to $W$, although the necessary coefficient of these terms does not appear
to have a natural explanation.

In the end, contrary to our initial expectations, 
the connection between area deficit and $W$ was not as simple as we anticipated.  
Nevertheless, it is our hope that the results and perspectives reported in this paper 
may 
 contribute to a deeper understanding of gravitational energy in the future.

Throughout, we use Greek indices $\alpha,\beta,\mu,\nu,\dots =0,1,\dots,d-1$ for spacetime RNC tensor expressions, while lowercase Latin indices $i,j,k,\dots =1,\dots d-1$ are spatial indices within the geodesic ball, and capital Latin indices $A,B,C,\dots =2,\dots,d-1$ are coordinate indices on the boundary surface of the ball, where $d$ is the spacetime dimension.

\paragraph{Note added:}  This paper (v2) includes corrections from the original version that 
was posted.  A summary of the changes made in v2 is included in appendix \ref{sec:corr}.

\section{Electric-magnetic decomposition of the Weyl tensor}\label{sec:EH}
Before presenting the calculations of the area and volume deficits, we briefly review the decomposition of the Weyl
tensor and its relation to the Bel-Robinson energy density $W$.  Given a spatial hypersurface with unit normal $u^\mu$,
the Weyl tensor $C_{\al\be\mu\nu}$ may be decomposed into its electric and magnetic parts, which are defined as 
spatial tensors on the hypersurface.  In adapted coordinates where the surface lies at $x^0=0$ and the unit normal is 
$u^\mu = \delta^\mu_0$, the electric-magnetic decomposition is given in terms of the following tensors:
\begin{align}
E_{ij} &= C_{0i0j} \label{eqn:Eij}\\
H_{ijk} &= C_{0ijk} \\
D_{ijkl} &= C_{ijkl} = F_{ijkl} +\frac{1}{d-3}(E_{ik}h_{jl}-E_{jk}h_{il}-E_{il}h_{jk}+E_{jl}h_{ik}),
\label{eqn:Dijkl}
\end{align}
where $h_{ij}$ is the spatial metric.
Up to dualization on anti-symmetric pairs of spatial indices, 
$E_{ij}$ is the electric-electric part, $H_{ijk}$ gives the electric-magnetic part  
and $D_{ijkl}$ gives the magnetic-magnetic part
\cite{Senovilla2000, Senovilla2001}. Under time reversal, $H_{ijk}$ changes sign, while 
$E_{ij}$ and $D_{ijkl}$ are invariant \cite{HOW2013}. Actually, 
since the Weyl tensor is traceless, the spatial trace of $D_{ijkl}$ contains $E_{ij}$:
\beq
h^{ij}D_{ikjl} = E_{kl},
\eeq
so that $F_{ijkl}$ defined in (\ref{eqn:Dijkl}) is spatially traceless. Note that $F_{ijkl}$
vanishes in $d=4$, in which case $D_{ijkl}$ is equivalent to $E_{ij}$, and $E_{ij}$ and $ B_{ij}\equiv\frac{1}{2}\epsilon_{jkl}H_i{}^{kl}$
are simply referred to as the electric and magnetic parts relative to $u^\mu$.

The Bel-Robinson super-energy density is given by the totally timelike component of the Bel-Robinson
tensor (\ref{eqn:Tarb}),  $W = T_{0000}$.  In terms of the electric-magnetic decomposition of the Weyl tensor, $W$ 
satisfies the simple relation \cite{Senovilla2000}
\beq \label{eqn:W}
W = \frac12\left[E^2+H^2+\frac14 D^2\right] = \frac12\left[\frac{d-2}{d-3} E^2 + H^2 + \frac14 F^2
  \right]
\eeq
where $E^2 = E_{ij} E^{ij} $, and similarly for $H$, $D$ and $F$. Observe that this can also be written, in any orthonormal basis containing $u^\mu$ as the timelike vector, as the sum of the squares of all components of the Weyl tensor
\beq
W=\frac{1}{8} \sum_{\al,\be,\mu,\nu =0}^{d-1} |C_{\al\be\mu\nu}|^2 .
\eeq
Note that $W$ is manifestly non-negative, and that $W=0$ if and only if the Weyl tensor vanishes, making it a good candidate for the small-sphere quasilocal energy in vacuum. The Bel-Robinson tensor is covariantly conserved in Einstein spaces, 
\beq
\nabla_\al T^\al{}_{\be\mu\nu} =0,
\eeq
and has a number of other important properties \cite{Senovilla2000} that were mentioned in the introduction. 

In $d=4$, using the 4-dimensional identity
$
C_{\al\gamma\rho\sigma}C^{\be\gamma\rho\sigma} =\frac{1}{4}\delta^\beta_\al C_{\mu\gamma\rho\sigma}C^{\mu\gamma\rho\sigma} 
$
the Bel-Robinson tensor (\ref{eqn:Tarb}) can be written in the more familiar forms
\begin{align}
T_{\al\be\mu\nu} & = C_{\al\rho\mu\sigma} C_\be{}^\rho{}_\nu{}^\sigma + 
C_{\al\rho\nu\sigma} C_\be{}^\rho{}_\mu{}^\sigma
-\frac{1}{8} g_{\al\be} g_{\mu\nu} C_{\zeta\rho\tau\sigma}C^{\zeta\rho\tau\sigma}\nonumber\\
& = C_{\al\rho\mu\sigma} C_\be{}^\rho{}_\nu{}^\sigma + *C_{\al\rho\mu\sigma} *C_\be{}^\rho{}_\nu{}^\sigma \label{eqn:Td=4} \hspace{1cm} \mbox{(only in $d=4$)}
\end{align}
where $*C$ is the Hodge dual of Weyl $C$, while 
the above expression (\ref{eqn:W}) reduces to the well-known formula $W=E^2+B^2$. The tensor (\ref{eqn:Td=4}) is fully symmetric and traceless. Actually, (\ref{eqn:Tarb}) is fully symmetric also in $d=5$, but not in higher $d$ \cite{Senovilla2000}. Notice, however, that $W$ ---and $p_\mu$--- is defined only from the fully symmetric part of $T_{\al\be\mu\nu}$ and thus we could always restrict ourselves to the fully symmetric part of (\ref{eqn:Tarb}).

In the following calculations,
we will look for contributions to the area and volume deficits proportional to $W$, in an attempt to relate these deficits
to gravitational energy.

\section{Variations of the area and volume}
We are interested in how the volumes and areas of small, geodesic balls in curved space differ from those in flat space.  A geodesic ball is constructed around a point 
$p$ by first choosing a timelike unit vector $u^\mu$ at $p$, and shooting out spacelike geodesics orthogonal to $u^\mu$.  
The ball is defined as the hypersurface generated by this congruence of spacelike geodesics, cut off at a fixed value $\ell$ of the affine parameter, which corresponds to the ball's radius.

We choose Riemannian normal coordinates (RNC) $\{x^\mu\}$ based at $p:= \{x^\mu =0\}$, so that the line-element at $p$ is the $d$-dimensional Minkowski metric $\eta_{\mu\nu} = \diag(-1,1,\dots,1)$. In RNC, the geodesics emanating from $p$ take the form $x^\mu =r n^\mu$ where $r$ is affine parameter and $n^\mu$ is the (unit) initial tangent vector at $p$. Adapting the coordinate system such that $u^\mu =\delta^\mu_0$, the spacelike geodesics generating the ball have initial tangent vectors $n^\mu =n^i \delta^\mu_i$
(hence $n^0|_p =0$). 
Then, along these geodesics their tangent vector fields are simply 
\beq
\frac{dx^\mu}{dr} = n^\mu = n^i \delta^\mu_i 
\eeq
on the entire geodesic ball. The ball is spacelike, its points having coordinates $x^\mu=n^\mu r =r n^i \delta^\mu_i$, so that it is locally defined by $t\equiv x^0=0$ and $\{x^i\}$ can then be used as coordinates on the ball. The unit normal to the ball $u_\mu$ is thus proportional $\nabla_\mu t$, and we also have $u_\mu n^\mu =0$ for all the above ``radial'' geodesics.

The deviations away from flatness near the point $p$ may be 
characterized by the curvature at $p$ via the standard RNC expansion
(see e.g.\ \cite{Brewin2009})
\beq
g_{\al\be}(x) = \eta_{\al\be}-\frac13 x^\mu x^\nu R_{\al\mu\be\nu} 
-\frac16 x^\mu x^\nu x^\rho\nabla_\mu R_{\al\nu\be\rho}
+ x^\mu x^\nu x^\rho x^\si\left(\frac2{45} R\indices{^\ga_\mu_\al_\nu} R_{\ga\rho\be\si}
-\frac1{20} \nabla_\mu \nabla_\nu
R_{\al\rho\be\si}\right), \label{eqn:metric}
\eeq
plus terms involving five or more powers of $x^\mu$.  Here, 
$R_{\al\mu\be\nu}$, $\nabla_\mu R_{\al\nu\be\rho}$, and 
$\nabla_\mu \nabla_\nu R_{\al\rho\be\si}$ are constants defined by the components of 
the curvature and its covariant derivatives 
at $p$.
Terms involving two factors of the curvature are needed since the first-order 
changes in area and volume are governed by the Ricci tensor \cite{Jacobson2015}, which
vanishes for the vacuum spacetimes that are the focus of this paper.  

In these coordinates, the induced metric $h_{ij}$ on the ball is 
\begin{align}
h_{ij} (x) =&\; \delta_{ij} -\frac13 x^k x^l R_{ikjl} - \frac16 x^k x^l x^m \nabla_k R_{iljm} \nonumber \\
&\;+ x^k x^l x^m x^n\left(-\frac2{45} R_{0kil}R_{0mjn} + \frac2{45} R\indices{^p_k_i_l}R_{pmjn} 
-\frac1{20} \nabla_k\nabla_l R_{imjn} \right) .\label{eqn:hij}
\end{align}
Note that this expression differs from the RNC expansion using the intrinsic hypersurface metric and 
curvature.  The Riemann tensors and covariant derivative $\nabla_k$ appearing in (\ref{eqn:hij})
correspond to the full spacetime metric $g_{\al\be}$, with components projected onto the 
hypersurface.  The reason for using the above form of the spatial metric is that it is easier to 
identify the contribution from the magnetic part of the Weyl tensor, appearing in the 
$R_{0kil}$ terms.  If instead one used the intrinsic RNC expansion, the magnetic Weyl terms 
would be hidden in the piece involving two intrinsic covariant derivatives of the intrinsic Riemann tensor.

The ball volume may be computed by integrating the volume form $\sqrt{h}dx^1\wedge \dots \wedge dx^{d-1}$ over the spatial ball.  
When the metric is expressed as a perturbation about a background, $h_{ij} = h^0_{ij} + 
\gamma_{ij}$, the volume density to second order in perturbations is
\beq
\sqrt{h} = \sqrt{h^0}\left(1+\frac12h_0^{ij}\gamma_{ij} + \frac18(h_0^{ij}\gamma_{ij})^2-\frac14
h_0^{ij} h_0^{kl}\gamma_{ik}\gamma_{jl} +\ldots\right).
\eeq
To integrate this quantity over the ball, it is convenient to use spherical coordinates $\{r,\theta^A\}$, 
where $\{\theta^A\}$ are coordinates on the $(d-2)$-sphere.  The $\sqrt{h^0}$ factor takes care
of the Jacobian when transforming to these coordinates, so the volume is given by
\begin{align}
V = \int d^{d-1} x\sqrt{h} = \int d\Omega_{d-2} \int_0^\ell dr\, r^{d-2}
\left(1+\frac12\delta^{ij}\gamma_{ij} + \frac18(\delta^{ij}\gamma_{ij})^2-\frac14
\delta^{ij} \delta^{kl}\gamma_{ik}\gamma_{jl} +\ldots\right) .\label{eqn:V}
\end{align}
The spatial metric perturbation $\gamma_{ij}$ defined by (\ref{eqn:hij}) 
involves terms with two, three, or four factors
of $x^k$. Observe that $x^k=r n^k$, where $n^k(\theta^A)$ describe the usual embedding of the unit $(d-2)$-sphere in Euclidean space ($\delta_{ij}n^i n^j =1$). When integrated over the sphere, only the spherically symmetric pieces of these factors 
survive.  This amounts to making the following replacements for the spherical integrals,
\begin{align}
&\int d\Omega_{d-2}\, x^k x^l = \frac{r^2\Omega_{d-2} }{d-1} \delta^{kl} ,\\
&\int d\Omega_{d-2}\, x^k x^l x^m = 0 ,\\
&\int d\Omega_{d-2}\, x^k x^l x^m x^n = \frac{r^4 \Omega_{d-2} }{d^2-1}(\delta^{kl}\delta^{mn} +
\delta^{km} \delta^{ln} + \delta^{kn}\delta^{lm})\equiv \frac{r^4\Omega_{d-2}}{d^2-1} \delta^{klmn},
\end{align}
where $\Omega_n = \frac{2\pi^{(n+1)/2}}{\Gamma(\frac{n+1}{2})}$ is the volume of the unit 
$n$-sphere.   

Using these replacements, we can immediately see that various parts of the $\delta^{ij}\gamma_{ij}$
term in (\ref{eqn:V}) vanish for a Ricci-flat metric
$
R_{\al\be}=0
$
that we assume throughout.  The two curvature corrections from the first 
line of (\ref{eqn:hij}) both vanish when integrated, the first being proportional to only Ricci tensor
components, and the second because $x^k x^l x^m$ integrates to zero over the ball.  The final
term in (\ref{eqn:hij}) involving two covariant derivatives of the Riemann tensor also vanishes,
because on using that the projector to the ball at $p$ is simply
\beq
h^{\al\be} \equiv \eta^{\al\be} +u^\al u^\be =\delta^{ij} \delta^\al_i \delta^\be _j , \hspace{1cm}
h_{\al\be}= \delta_{ij} \delta_\al^i \delta_\be ^j , 
\eeq
one easily gets
\beq
\delta^{ij}\delta^{klmn}\nabla_k \nabla_l R_{imjn} = \delta^{klmn}\nabla_k\nabla_l
 R_{mn}
+ u^\al u^\be(h^{\mu\nu}\eta^{\rho\si} +\eta^{\mu\rho}\eta^{\nu\si} + \eta^{\mu\si}\eta^{\nu\rho})
\nabla_\mu\nabla_\nu R_{\al\rho\be\si}, \label{nablanablaterm}
\eeq
and by the Bianchi identity,
\beq
\nabla^\nu R_{\al\rho\be\nu} = \nabla_\rho R_{\al\be}-\nabla_\al R_{\rho\be},
\eeq
we see that all of the above terms are constructed solely from derivatives of the Ricci 
tensor.  

The remaining terms are all quadratic in the Weyl tensor, and arise from the first two terms
in the second line of (\ref{eqn:hij}), as well as from the terms in (\ref{eqn:V}) that are quadratic
in $\gamma_{ij}$. 
Using the electric-magnetic decomposition of the Weyl tensor from equations (\ref{eqn:Eij})-(\ref{eqn:Dijkl}),
the curvature squared contributions to the volume  (\ref{eqn:V})
after performing the angular and radial integrals become
\begin{align}
\delta V &= \frac{\Omega_{d-2} \ell^{d+3}}{9(d^2-1)(d+3)}\delta^{klmn}\left[-\frac15 H\indices{_k^i_l}
H_{min}+\left(\frac15-\frac14\right) D\indices{^p_k^i_l}D_{pmin} +\frac18 E_{kl} E_{mn} \right]\nonumber\\
&=\frac{\Omega_{d-2} \ell^{d+3}}{15(d^2-1)(d+3)}\left[ -\frac{D^2}{8} - \frac{H^2}{2} +\frac{E^2}{3}
 \right].\label{eqn:deltaV}
\end{align}

We can obtain the change in area from this expression due to the following observation.  The radial
coordinate $r = \sqrt{x^k x^l \delta_{kl}}$ foliates the ball by surfaces orthogonal to the geodesics emanating
from the center.  The unit normal to constant $r$ surfaces is equal to the covariant tangent vector
for the geodesics, $n_a = \nabla_a r$, and hence the volume form $\eta$ of the hypersurface 
is related
to the area form $\mu$ on the constant $r$ surfaces via
\beq
\eta = n\wedge \mu.
\eeq
This means our volume integral may be expressed as 
\beq
V(\ell) = \int_0^\ell dr\, A(r),
\eeq
so that $A(\ell) = \frac{\partial}{\partial \ell} V(\ell)$.  Applying this formula to equation (\ref{eqn:deltaV})
immediately gives
\beq \label{eqn:dA}
\delta A = \frac{\Omega_{d-2} \ell^{d+2}}{15(d^2-1)}\left[-\frac{D^2}{8} -\frac{H^2}{2} +\frac{E^2}{3}\right].
\eeq

From the calculation in non-vacuum spacetimes \cite{Jacobson2015}, where $\delta A$ is proportional to $-G_{00}$ ($G_{\al\be}$ being the Einstein tensor), and thereby, via the Einstein field equations, proportional to $-T_{00}$ ($T_{\al\be}$ being the energy-momentum tensor), we were expecting several obvious properties for $\delta A$: it should be proportional to the timelike component of a tensor field, and it should have sign properties. For instance, one would expect it to be negative definite, and this is true in $d=4$, but not in general for other dimensions $d$. But even in $d=4$, where the above expression reduces to
\beq \label{eqn:dA(d=4)}
\delta A = \frac{\Omega_{2} \ell^{6}}{225}\left[ -B^2-\frac{E^2}{6}\right] \hspace{1cm} \mbox{(only in $d=4$)},
\eeq
the area deficit fails to satisfy all the desired properties of the quasilocal energy listed in the introduction.  
In particular, although it is proportional to the timelike component of the four-index tensor
\beq
t_{\al\be\mu\nu} = C_{\al\rho\mu\sigma} C_\be{}^\rho{}_\nu{}^\sigma + 6 *C_{\al\rho\mu\sigma} *C_\be{}^\rho{}_\nu{}^\sigma,
\eeq
the factor of $6$ in the second term makes it impossible that the momentum vector $p_\al = - t_{\al\be\mu\nu}u^\be u^\mu u^\nu$ be future pointing for generic $u^\mu$ \cite{Senovilla2000}. Choose for instance a case (in an orthonormal basis including $u^\mu =\delta^\mu_0$) with $E_{1i}=B_{1i}=0$, $E_{22}=-E_{33}=-7B_{23}/2$, and $B_{22}=-B_{33}=2E_{23}/7$, then contracting $p_\mu$ with the future null vector $\delta^\mu_0+\delta^\mu_1$ one gets $25 B^2/4$ which is strictly positive, proving that $p_\mu$ is not future pointing. This failure occurs to any tensor with different weights between the $C^2$ and the $*C^2$ summands.
The only tensor (up to prefactors) quadratic in the Weyl tensor and having the required dominant property is 
the generalized Bel-Robinson tensor, given in arbitrary spacetime dimension by equation (\ref{eqn:Tarb}) 
\cite{Senovilla2000}. 
Its total timelike component, the Bel-Robinson super-energy density $W=T_{0000}$, may be expressed in terms of the electric and magnetic parts of the Weyl tensor according to (\ref{eqn:W}).

There is no unique way to write our result (\ref{eqn:dA}) in terms of $W$; some possibilities would be 
\beq \label{eqn:dV}
\delta A = \frac{\Omega_{d-2} \ell^{d+2}}{15 (d^2-1)}\left[-W+\frac{5 }{6} E^2\right]=
\frac{2\Omega_{d-2} \ell^{d+2}}{45 (d^2-1)}\left[W-\frac{5 }{4} H^2-\frac{5}{16} D^2 \right]
\eeq
 but there are of course many others.
 This is a little puzzling, as the expected result (something proportional to $W$) is not what arises at first. 

At this point, it is necessary to remark that we are actually trying to compare two fully different spacetimes locally---a generally curved one with flat spacetime---and this is intricate. One cannot be certain of what is exactly the analogue of a given flat-spacetime geodesic ball in the curved spacetime. 
For instance, 
in the first order calculation in the presence of matter, the natural prescription seems to be to vary the geodesic radius
of the ball such that the volume is held fixed \cite{Jacobson2015}. 
At second order, there is 
the additional possibility of varying the shape of the ball in a way that depends on the local
gravitational field (the first order contribution of these shape deformations to the area vanishes).  
These possibilities will be analyzed in what follows, 
with a goal of finding a prescription that yields
an area variation proportional to $W$, 
in accordance with the criteria presented in the introduction for quasilocal gravitational energy.

\section{Ball deformations} \label{sec:deformations}
We now allow for variations of the ball radius and compute the volume and area variations under these circumstances. Instead of formula (\ref{eqn:V}) we now have
\begin{align}
V = \int d\Omega_{d-2} \int_0^{\ell +\delta\ell}  dr\, r^{d-2}
\left(1+\frac12\delta^{ij}\gamma_{ij} + \frac18(\delta^{ij}\gamma_{ij})^2-\frac14
\delta^{ij} \delta^{kl}\gamma_{ik}\gamma_{jl} +\ldots\right) \label{eqn:V2}
\end{align}
where $\delta \ell$ is the deformation of the ball radius which we assume may depend on the direction, taking the following form 
\beq\label{eqn:deltaL}
\delta\ell = \underbrace{\delta\ell_1 +\tilde\delta\ell_1}_{O(1)} +\underbrace{X}_{O(2)}.
\eeq
Here $\delta\ell_1$ and $X$ are the spherically symmetric pieces of the first- and second-order perturbations, respectively, while $\tilde\delta\ell_1$ is the part of the first-order perturbation depending on the direction. It was proven in \cite{Jacobson2015} that $\delta\ell_1$ can be chosen to keep the volume fixed at first order (the same as the volume in flat space). In the Ricci flat case such a choice requires $\delta\ell_1 =0$, that we assume from now on unless otherwise stated. Then, (\ref{eqn:V2}) becomes
\beq
V=V^{\flat}(\ell) +\delta V + \Delta V
\eeq
where $V^{\flat}(\ell)$ denotes the volume of a radius $\ell$ round ball in flat space, $\delta  V$ is 
the volume variation purely due to curvature, given in (\ref{eqn:deltaV}), and
\begin{align}
\Delta V &= \ell^{d-2} \int d\Omega_{d-2} \left(X +\frac{1}{2} \delta^{ij} \gamma_{ij} \tilde\delta\ell_1 +\frac{d-2}{2\ell}\, (\tilde\delta\ell_1)^2\right) \nonumber \\
& =  \ell^{d-2} \left[\Omega_{d-2} X +\int d\Omega_{d-2} \left(\frac{d-2}{2\ell}\, (\tilde\delta\ell_1)^2-\frac{\ell^2}{6}E_{ij}n^i n^j \,  \tilde\delta\ell_1 \right) \right].\label{eqn:DeltaV}
\end{align}
As a function defined on the  $(d-2)$-sphere,
$\tilde\delta\ell_1$ can be expanded in spherical harmonics. Letting $s$ denote the ``spin,'' we have
\beq
\tilde\delta\ell_1 = \sum_{s=1}^\infty Y_{i_1\dots i_s} n^{i_1}\dots n^{i_s} 
\eeq
where $Y_{i_1\dots i_s}$ are totally symmetric and traceless for all $s>1$. 
Formula (\ref{eqn:DeltaV}) can then be explicitly computed. The second term in (\ref{eqn:DeltaV}) contributes like
\beq
\Omega_{d-2} \ell^{d-3} (d-2)  \sum_{s=1}^\infty c_s Y_{[s]}^2 
\eeq
where $Y_{[s]}^2\equiv Y_{i_1\dots i_s}Y^{i_1\dots i_s}$ and $c_s$ are some constant factors---depending on $s$ and the dimension $d$---whose explicit expression will not be needed in what follows except for 
\beq \label{eqn:c2}
c_2=\frac{1}{d^2-1}. 
\eeq
Concerning the last term in (\ref{eqn:DeltaV}), all the integrals evaluate to zero (due to either an odd number of $n$'s or to the tracelessness of $E_{ij}$ and $Y_{i_1\dots i_s}$) except for the spin-2 piece of the deformation that couples to $E_{ij}$ to give
\beq
-\Omega_{d-2} \ell^{d} \frac{1}{3(d^2-1)} Y^{ij} E_{ij} .
\eeq
Putting everything together we arrive at
\beq
\Delta V = \Omega_{d-2} \ell^{d-3}  \left[ X\ell +(d-2)\sum_{s=1}^\infty c_s Y_{[s]}^2 -\frac{\ell^3}{3(d^2-1)} Y^{ij}E_{ij}  \right] \, . 
\eeq
so that 
\beq
V=V^\flat (\ell) +\delta V+ \Omega_{d-2} \ell^{d-3}  \left[ X\ell +(d-2)\sum_{s=1}^\infty c_s Y_{[s]}^2 -\frac{\ell^3}{3(d^2-1)} Y^{ij}E_{ij}  \right] \label{eqn:totalV}
\eeq
with $\delta V$ given in (\ref{eqn:deltaV}).

In order to compute the area of the surface
limiting the deformed geodesic ball, we note that the embedding of such a surface into the hypersurface $x^0=0$ containing the ball is given by
\beq\label{eqn:embed}
x^i = n^i \left(\ell +\tilde\delta\ell_1 + X \right)
\eeq
where $n^i(\theta^A)$ were introduced after Eq.(\ref{eqn:V}) as functions of the intrinsic coordinates $\{\theta^A\}$ in the $(d-2)$-sphere. Taking into account that
\beq
\delta_{ij} n^i \frac{\partial n^j}{\partial \theta^A}=0, \hspace{1cm} \delta_{ij}\frac{\partial n^i}{\partial \theta^A}\frac{\partial n^j}{\partial \theta^B} = \Omega_{AB}
\eeq
where $\Omega_{AB}$ is the standard metric on the round $(d-2)$-sphere, a straightforward calculation using (\ref{eqn:hij}) gives, for the first fundamental form $q_{AB}$ inherited on the limiting surface,
\begin{align}
q_{AB} & = \underbrace{\ell^2 \Omega_{AB}}_{q^0_{AB}} +\underbrace{\left[2\ell\,  \tilde\delta\ell_1\Omega_{AB} -\frac{\ell^4}{3} n^k n^l D_{ikjl} \frac{\partial n^i}{\partial \theta^A}\frac{\partial n^j}{\partial \theta^B}\right]}_{\delta q^1_{AB}}-  \underbrace{\frac{1}{6}
\ell^5 n^k n^l n^m \nabla_k R_{iljm}\frac{\partial n^i}{\partial \theta^A}\frac{\partial n^j}{\partial \theta^B}}_{\delta q^{3/2}_{AB}}  \nonumber \\
& +\underbrace{ \Omega_{AB}\left[2\ell X + (\tilde\delta\ell_1 )^2 \right]+\frac{\partial n^i}{\partial \theta^A}\frac{\partial n^j}{\partial \theta^B}\left(
\sum_{s,\hat s =1}^\infty s \hat s Y_{i i_2\dots i_s}Y_{jj_2\dots j_{\hat s}}n^{i_2}\dots n^{i_s} n^{j_2}\dots n^{j_{\hat s}} - \frac{4\ell^3}{3}\tilde\delta\ell_1 n^k n^l D_{ikjl}  \right)}_{\delta q^2_{AB}}\nonumber \\
&+ \underbrace{\ell^6 n^k n^l n^m n^p\left(-\frac{2}{45} H_{kil} H_{mjp}+\frac{2}{45} D^q{}_{kil}D_{qmjp} -\frac{1}{20} \nabla_k\nabla_l R_{imjp} \right)\frac{\partial n^i}{\partial \theta^A}\frac{\partial n^j}{\partial \theta^B}}_{\delta q^2_{AB}}.\label{eqn:1FF}
\end{align}
Having chosen vanishing volume variation at first order, the first-order variation of area is also vanishing; furthermore, the term $q_0^{AB} \delta q^{3/2}_{AB}$ will not contribute upon integration, because 
\beq
q_0^{AB} \left(\ell \frac{\partial n^i}{\partial \theta^A}\right)\left(\ell \frac{\partial n^j}{\partial \theta^B}\right) = \delta^{ij} -n^i n^j 
\eeq
so that this term contains an odd number of $n$'s. We  can thus concentrate on the second-order variation of area. The formula for this is 
\beq
\frac{\ell^{d-2}}{2} \int d\Omega_{d-2} \left(q_0^{AB}\delta q^2_{AB} +\frac{1}{4} (q_0^{AB}\delta q^1_{AB})^2 -\frac{1}{2} q_0^{AC} q_0^{BD} \delta q^1_{AB} \delta q^1_{CD}\right)
\eeq
so that a somewhat long but direct calculation leads to
\beq
A =A^{\flat}(\ell) +\delta A + \Delta A,
\eeq
where $A^{\flat}(\ell)$ is the area of a radius $\ell$ round sphere in flat space, $\delta A$  given in (\ref{eqn:dA}) involves curvature squared terms, and
\beq \label{eqn:DA}
\Delta A = \Omega_{d-2} \ell^{d-4}\left[X \ell (d-2) +\sum_{s=1}^\infty b_s Y_{[s]}^2-\frac{\ell^3 d}{3(d^2-1)} Y^{ij} E_{ij} \right]
\eeq 
where $b_s$ are constant factors ---depending on $s$ and $d$--- whose explicit expression is unimportant for our purposes excepting 
\beq\label{eqn:b2}
b_2=\frac{d^2-3d+4}{d^2-1}.
\eeq 

Here we note that of all the shape deformations parameterized by $Y_{[s]}$, only the spin-2 
deformation gives a different contribution to the area in curved space than in
flat space.  This difference is given by the $Y^{ij} E_{ij}$ term in (\ref{eqn:DA}).  In fact, it is only
the component of $Y_{ij}$ that is aligned with $E_{ij}$ that contributes differently than in 
flat space.  To see this in more detail, note that the set of spatial (relative to $u^\mu$) 2-covariant symmetric and trace-free tensors at $p$ is a vector space of dimension $(d+1)(d-2)/2$ with a natural positive definite scalar product given by $Y_{ij}\hat{Y}^{ij}$. Thus, given $E_{ij}$ as data, $Y_{ij}$ will have a component along $E_{ij}$ and another part $Z_{ij}$ orthogonal to $E_{ij}$:
\begin{equation}
Y_{ij}=\ell^3\left( \gamma E_{ij}+ Z_{ij} \right). \label{YE}
\end{equation}
Here, $Z_{ij}$ is symmetric, traceless and orthogonal to $E_{ij}$ ($E_{ij}Z^{ij}=0$) while $\gamma$ is a shorthand for $Y_{ij}E^{ij}/(\ell^3 E^2)$ .
Since all shape deformations with $s\neq 2$ and the component of $Y_{ij}$ orthogonal 
to $E_{ij}$ make the same contribution to the area in flat 
space  as in any curved space, they cannot be fixed  in terms of the local 
gravitational field at this order in perturbations.  We therefore assume that the only shape 
deformation is given by $Y_{ij}$ aligned with $E_{ij}$.  

With this in mind, we can rewrite the area of ball's boundary as
\beq
A= A^\flat(\ell) + \frac{\Omega_{d-2}\ell^{d+2}}{(d^2-1)} \left[-\frac{W}{15} +E^2\left(\gamma^2(d^2-3d+4)-\frac{\gamma d}{3}  +\frac{1}{18} \right)+ X  
\frac{(d-2)(d^2-1)}{\ell^5}\right] \label{eqn:difA} .
\eeq
The remaining freedom encoded in $\gamma$ and $X$ is obviously enough to get something proportional to $W$, and generically the radius variation $X$ has to be nonzero for this to occur.  
This is because the coefficient multiplying the $E^2$ term has a minimum at $\gamma_0 = 
d/[6(d^2-3d+4)]$, in which case the coefficient becomes $(d-2)(d-4)/[36(d^2-3d+4)]$, which is 
positive for $d>4$.  Oddly enough, precisely when $d=4$ and  $\gamma = \gamma_0 = 1/12$,
we find that this coefficient vanishes, leaving only $-W/15$ in the area variation if the 
average radius is 
held constant by setting $X=0$.  
As shown in section \ref{subsec:K}, this value of $\gamma$ arises when requiring that the 
trace of the second fundamental form of the ball remain constant to first order 
in the RNC expansion.   
However, in the higher dimensional case, there must be a spherically symmetric 
radius variation resulting in a nonzero value of $X$ in order to cancel the $E^2$ terms, and 
this, along with the shape deformation, should be determined by independent arguments.  
There are several routes that can be pursued, with no clear preference for one in principle. We discuss this in the following sections.

\section{Fixing the deformation by independent arguments} \label{sec:fixing}
As seen in formula (\ref{eqn:difA}), there are two parameters to be fixed by independent arguments: the ``amount'' of spin-2 deformation along the electric-electric part of the Weyl curvature, encoded in $\gamma$; and the total spherically symmetric size of the ball, encoded in the second-order variation of its radius $X$. We start with the former, which is the more intriguing one, and takes care of the non-isotropic, basically quadrupolar, nature of the gravitational field at the center of the ball $p$. 

\subsection{Causal diamond deformations}\label{subsec:diamond}

A particularly natural way to define the ball deformation 
is to choose its shape to 
ensure that the boundary of its causal development takes the form of two intersecting cones,
and does not develop caustics except at the apexes of the future and past cones.  This statement
of course is perturbative in the curvature at the center of the ball $p$, and we will only need to work to first order in the 
curvature at $p$ to see the first effect of the shape deformation. We continue to assume that the spacetime geometry is Ricci flat.  

The strategy is to follow the timelike geodesic defining the frame at the center of the ball
a fixed proper time to the future and past to define the apexes of the cones, 
and then send out null geodesics to define the cone null hypersurface.  A straightforward way to find such a 
hypersurface is to assume these null geodesics are simply the gradient of a scalar, $k_\mu = \nabla_\mu \phi$,
and then we need only impose that $\nabla_\mu \phi$ is null to ensure that $k^\mu$ is tangent to affinely 
parameterized geodesics.  

We start with the future null cone.  In flat space, this surface is defined by $t+r=\ell$ for a ball
of radius $\ell$.  So we take $\phi = t+r+\mathcal{O}(R)$.  The first order in curvature correction to 
$\phi$ we could guess needs to be of the form $\beta(r,t) x^i x^j E_{ij}$, since there are no other 
terms that can be formed that are linear in the electric/magnetic parts of the Weyl tensor where 
all the indices are contracted with $x^k$.  Another way to see that this is the correct type of term to 
add is to calculate how the norm of $\nabla_\mu(t+r)$ changes at first order in the RNC expansion.  
Using that $\nabla_\mu r =\frac{1}{r} \delta_{\mu j}x^j$
we have
\begin{align}
g^{\alpha\beta} \nabla_\alpha (t+r) \nabla_\beta (t+r) &= (\delta_\alpha^0 +\frac{1}{r} \delta_{\alpha i}x^i)(\delta_\beta^0+\frac{1}{r} \delta_{\beta j}x^j)
(\frac13 x^\gamma x^\delta R\indices{^\alpha_\gamma^\beta_\delta}) \nonumber\\
&=\frac13 x^i x^jE_{ij}\left(1+2\frac{t}{r}+\frac{t^2}{r^2}\right),\label{eqn:null1}
\end{align}
so it is clear that we should be able to cancel this by choosing $\phi =t + r +\beta x^i x^j E_{ij}$.
The norm of $\phi$ is corrected by a contribution
\beq
2\nabla_\alpha(\beta x^i x^j E_{ij}) (\nabla_\beta t +\nabla_\beta r) \eta^{\alpha\beta} = x^i x^j E_{ij}
\left(-2\partial_t \beta + 2\partial_r\beta + \frac{4\beta}{r}\right). \label{eqn:null2}
\eeq
The sum of (\ref{eqn:null1}) and (\ref{eqn:null2}) must vanish for $\nabla_\mu\phi$ to be null at this order in RNC.  
Switching to null coordinates $u = t-r$, 
$v=t+r$, the condition becomes
\beq
\partial_u\beta + \frac{2\beta}{u-v} -\frac{v^2}{3(u-v)^2} = 0.
\eeq
This has the general solution
\beq
(u-v)^2\beta = \frac{v^2 u}{3} + C(v) \hspace{3mm} \Longrightarrow \hspace{3mm}
\phi = t+r +\frac{1}{4}\left(\frac{1}{3} (t-r) (t+r)^2 +C(t+r)\right) E_{ij} n^i n^j .
\eeq
To fix the arbitrary function $C$, we require that $\phi$ coincides with $t$ as $r\rightarrow 0$.  
This immediately implies
\beq
C(v) = \frac{-v^3}{3}, 
\eeq
and hence 
\beq
\phi =t+r - \frac{1}{6} r (t+r)^2 E_{ij} n^i n^j .\label{eqn:phi}
\eeq

An analogous calculation can be done for the past null cone, which turns out to be a level set of the function
\beq
\psi =t-r+\frac{1}{6} r (t-r)^2 E_{ij} n^i n^j .\label{eqn:psi}
\eeq

Our ball should lie at the intersection of the two null hypersurfaces defined by $\phi=\ell$, $\psi=-\ell$, where here $\ell$ can be seen as the proper time elapsed between $p$ and the apexes of the cones along the timelike geodesic tangent to $u^\mu$.
Hence we should look for the values of $t$ and $r$ that solve the equations
\begin{align}
t+r - \frac{1}{6} r (t+r)^2 E_{ij} n^i n^j  &= \ell \label{eqn:phi=l} \\
t-r+\frac{1}{6} r (t-r)^2 E_{ij} n^i n^j &= -\ell.  \label{eqn:psi=-l}
\end{align}
Adding these two gives 
\beq
t \left(1-\frac{1}{3} x^i x^j E_{ij}\right)=0 
\eeq
which is solved by $t=0$. Hence, at this order in the expansion, the intersection of the two null cones lies on the chosen spatial hypersurface $t=0$; there is no bending in 
the time direction. 
Concerning the coordinate $r$, we set $t=0$ in either of (\ref{eqn:phi=l}) or (\ref{eqn:psi=-l}) to get
\beq
r = \ell +\frac{1}{6}r^3 n^i n^j E_{ij},
\eeq
and, since we are working only to first order in $E_{ij}$, this gives the solution for $r$ as 
\beq
r = \ell\left(1+\frac16 \ell^2 n^i n^j E_{ij}\right),
\eeq
or equivalently the shape deformation
\beq
\tilde\delta \ell_1 = n^i n^j Y_{ij} = \frac{1}{6} \ell^3 n^i n^j E_{ij} .
\eeq
This implies that all non-spin-2 deformations vanish and sets $\gamma =1/6$ and $Z_{ij}=0$ in (\ref{YE}).  Therefore, if our geodesic spatial ball is to be the base of a small causal diamond, its shape in a vacuum spacetime is fully specified by the electric-electric part of the Weyl tensor, in such a way that the spin-2 deformation $Y_{ij}$ must be aligned with $E_{ij}$ a fixed amount $\gamma =1/6$. 

This procedure can be carried out to higher order in the curvature expansion.  However, the only 
effect on the area due to the next order shape deformation will come from the overall change in 
radius that they produce, and hence is degenerate with the $X$ deformation.  This is actually just an ambiguity in how to define the size of the ball, and
could be compensated by changing the value $\ell$ of the proper time corresponding to the apexes of the cones.  

\subsection{Using the geometry of the ball boundary}\label{subsec:K}
The above prescription for choosing the shape of the ball turns out to be equivalent to imposing
a condition on the extrinsic geometry of the boundary.  
We need to compute the second fundamental form characterizing this extrinsic geometry.
Using spherical coordinates $\{r,\theta^A\}$ based at the center $p$ the embedding (\ref{eqn:embed}) into the ball reads simply as 
\beq
r=\ell +{\sum_{s=1}^\infty }Y_{i_1\dots i_s}n^{i_1} \dots n^{i_s} + X,
\eeq
hence a normal to the boundary is
\beq
N=dr -\left(\sum_{s=1}^\infty s\, Y_{i_1\dots i_s}n^{i_2} \dots n^{i_s}  \right)\frac{\partial n^{i_1}}{\partial \theta^A} d\theta^A
\eeq
which is of unit length at the required order, while the tangent vector fields are
\beq
\vec e_A =\left(\sum_{s=1}^\infty s\, Y_{i_1\dots i_s}n^{i_2} \dots n^{i_s}  \right)\frac{\partial n^{i_1}}{\partial \theta^A}  \partial_r +\partial_{\theta^A}\, .
\eeq
The second fundamental form can be computed from the formula
\beq
K_{AB} = -N_i e^j_A \tilde\nabla_j e^i_B
\eeq
where $\tilde\nabla$ is the covariant derivative on the hypersurface $x^0=0$. Using that 
\beq \label{eqn:Gamma0}
\stackrel{0}{\Gamma^r_{Ar}} =0, \hspace{3mm} \stackrel{1}{\Gamma^r_{AB}}=\frac{2}{3} r^3 n^k n^l \frac{\partial n^i}{\partial\theta^A} \frac{\partial n^j}{\partial\theta^B} D_{ikjl} {-\tilde{\delta}\ell_1 \Omega_{AB}}
\eeq
and that the {\em functions} $n^i(\theta^A)$ satisfy on the unit round sphere (here $\overline\nabla$ in the covariant derivative on the unit sphere)
\beq
\overline\nabla_A \overline\nabla_B n^i = -\Omega_{AB} n^i
\eeq
we get for the second fundamental form the following expression including terms linear in the curvature (keeping the Ricci flat condition)
\begin{align}
K_{AB} =&\;
\left(\ell + \sum_{s=1}^\infty \az{(s+1)}\, Y_{i_1\dots i_s}n^{i_1} \dots n^{i_s} \right) \Omega_{AB}  
\nonumber \\
&\;-\left(\sum_{s=2}^\infty s(s-1) Y_{iji_3\dots i_s}n^{i_3}\dots n^{i_s}
+\frac{2}{3} \ell^3 n^k n^l D_{ikjl} \right) \frac{\partial n^i}{\partial\theta^A} \frac{\partial n^j}{\partial\theta^B} \, .\label{eqn:2FF}
\end{align}

In particular, its trace $K=q^{AB}K_{AB}$ reads at this order
\beq
K = \frac{d-2}{\ell}+\frac{1}{\ell^2} \sum_{s=1}^\infty \az{(d-2+s)(s-1)} Y_{i_1\dots i_s}n^{i_1} \dots n^{i_s} -\frac{\ell}{3} E_{ij}n^i n^j  \label{eqn:K}
\eeq
which can be conveniently rewritten as
\beq
K = \frac{d-2}{\ell}+\frac{1}{\ell^2} \sum_{s\neq 2} \az{(d-2+s)(s-1)} Y_{i_1\dots i_s}n^{i_1} \dots n^{i_s}+n^i n^j\left( \az{\frac{d}{\ell^2}} Y_{ij} -\frac{\ell}{3} E_{ij}\right).
\eeq
In other words, the trace of the second fundamental form of the boundary within the hypersurface $t=0$ is affected by the gravitational field only through a term proportional to $E_{ij}n^i n^j$. 
This trace is constant on the entire surface only if all non-spin-2 deformations vanish and the spin-2 deformation is aligned with the electric-electric part
\beq \label{eqn:YK}
Y_{ij} = \az{ \frac{\ell^3}{3d}} E_{ij} \, .
\eeq
Observe that the requirement $K=$ constant turns out to be equivalent to keeping $K$ stationary with respect to its flat-space sphere value at this order.

\subsection{Null expansions on the ball's boundary}\label{subsec:nulltheta}
An equivalent way of defining the ball's boundary shape is to keep its  null expansions constant on the entire boundary. This is again equivalent to keeping them stationary with respect to flat spacetime. In order to see this, we compute the null expansions on the $(d-2)$-dimensional deformed sphere which bounds the geodesic ball. Given that we already know the trace $K$ given in (\ref{eqn:K}), we only need to know the second fundamental form of the hypersurface $t=0$ around $p$. This can be obtained by using that the unit normal is $u_\mu =-\delta_\mu^0$ as (at first non-trivial order in vacuum)
\beq
K_{ij} =\Gamma^0_{ij}= \frac{1}{3} x^\delta \left(R^0{}_{i\delta j} + R^0{}_{j\delta i} \right)=
\frac{1}{3} \left(H_{ijk} + H_{jik} \right)x^k .
\eeq
This could also be easily derived by using the Gauss and Codazzi equations of the ball. 

On the ball's boundary at this order we must use $x^k=\ell n^k$, and pulling back $K_{ij}$ we obtain the second fundamental form $\varkappa_{AB}$ of the boundary with respect to the timelike future-pointing unit normal $-dt$. This reads
\beq
\varkappa_{AB} =\frac{\ell}{3} \left(H_{ijk} + H_{jik} \right)n^k \frac{\partial n^i}{\partial\theta^A} \frac{\partial n^j}{\partial\theta^B} .\label{eqn:kappa}
\eeq
Its trace $\varkappa \equiv q^{AB}\varkappa_{AB}$ therefore vanishes at this order
\beq
\varkappa = q_0^{AB}\varkappa_{AB} =\frac{1}{\ell^2} \Omega^{AB}\varkappa_{AB} =\frac{1}{3\ell} \left(H_{ijk} + H_{jik} \right)n^k (\delta^{ij} - n^i n^j) =0 .\label{eqn:kappatrace}
\eeq
The two future null normals to the boundary are $k_\pm\equiv -dt \pm N$ and thus the null expansions are simply given by $\theta_\pm =\varkappa \pm K = \pm K$. In conclusion, at this order in the RNC expansion for vacuum spacetimes, these null expansions are constant over the whole ball's boundary  if, and only if,  $K$ is, and this was previously demonstrated to hold only if the ball is deformed in a spin-2 manner exclusively, aligned with $E_{ij}$ an amount given by \az{$\gamma =1/(3d)$}. In that case, the null expansions
coincide with their flat spacetime analogues too.

\section{Volume control}\label{sec:volume}
In the presence of matter, the first law of causal diamonds relates the area variation to the energy density
inside the ball when the Einstein field equations hold \cite{Jacobson2015}.  
This proportionality between area and energy density variations holds for arbitrary linearized perturbations of 
a finite-sized ball in flat space,
provided that the volume of the perturbed ball is the same as in flat space.  
This provides a motivation for considering variations at fixed volume at first order.  
A second motivation comes from considering small balls in an arbitrary curved spacetime.
In this small ball limit, one may interpret the curvature terms in the RNC expansion (\ref{eqn:metric}) 
as a perturbation of the locally flat metric 
at the center of the ball.
In this case, both the area and volume variations are determined by the 
 Einstein tensor component $G_{00}$, so that the area deficit remains proportional to the energy density
even if the volume fluctuates, albeit with a proportionality factor depending 
on the volume variation.  However, when the first law is interpreted as a maximality
condition for the entanglement entropy of the ball, the constant of proportionality is meaningful, since it determines the 
Newton constant $G$ appearing in the field equations and the 
Bekenstein-Hawking entropy, $S_{\text{BH}} = A/4\hbar G$.  Since this latter quantity is closely related to area terms in
entanglement entropy, the entire maximal entanglement picture is consistent only when the volume is held fixed
\cite{Jacobson2015}.

The above arguments for holding the volume fixed apply only at first order in perturbations away from flat space.  
Here, we will investigate the area variation holding the volume fixed at second order, and 
find that for no choice of the shape deformation $Y_{ij}$ do we obtain an answer proportional
to $W$.  
This motivates us to consider other prescriptions. 
One possibility is that the ball in flat space to which we should compare does not correspond
to the round sphere, and we discuss how a flat space comparison ball can be chosen to produce 
the desired answer.  Unfortunately, this procedure is somewhat {\it ad hoc}, and 
we leave the problem of finding a better justification for this prescription to future work.

Keeping only the $E_{ij}$-aligned, spin-2 shape deformation  $Y_{ij} = \gamma \ell^3 E_{ij}$, 
the total volume (\ref{eqn:totalV}) of the ball reads
\beq
V = V^\flat(\ell) +\frac{\Omega_{d-2} \ell^{d+3}}{d^2-1}\left[-\frac{W}{15(d+3)} + 
E^2\left(\gamma^2(d-2)-\frac{\gamma}{3} +\frac1{18(d+3)}\right) \right] + \Omega_{d-2} \ell^{d-2}
X.
\eeq
Choosing $X$ such that $V-V^\flat(\ell)$ vanishes, and substituting into equation (\ref{eqn:difA})
for the area variation, we find the expression for the area variation at fixed volume to second order,
\beq\label{eqn:dAV}
\delta A\big|_V = \frac{\Omega_{d-2}\ell^{d+2}}{3(d+3)(d^2-1)}\left[-W+E^2\left(3d(d+3)\gamma^2-2(d+3)\gamma +\frac56\right) \right].
\eeq
The coefficient of the $E^2$ takes its minimal value of $(d-2)/2d>0$ when the shape 
deformation is chosen with $\gamma = \gamma_m= 1/3d$. Hence, for no choice of shape
deformation will the area change at fixed volume be proportional to $W$.  

There are some remaining options to consider.  First, we could relax the condition of fixing the 
volume of the ball, and merely choose $X$ to ensure the area variation is proportional to $W$.  
This could still be consistent with the first law of causal diamonds, because we are considering 
second order variations, whereas the fixed volume constraint was only derived at first order.  
A downside of this approach is that we can obtain any coefficient of $W$ 
in the area variation, simply by choosing $X$ to be (here $\alpha$ is an arbitrary constant)
\beq
X = \frac{\ell^5}{(d-2)(d^2-1)}\left[ -E^2\left(\gamma^2(d^2-3d+4)-\frac{\gamma d}{3}+\frac1{18}\right) +\alpha W\right].
\eeq
The ability to shift $X$ by an arbitrary amount proportional to $W$ leads to a similar ambiguity
in the area variation, and underlies the need to find an independent justification for 
fixing the overall size of the ball.

One way to remedy this overall radius problem is to consider variations of well-defined geometric quantities that are equivalent to varying the area while keeping something fixed (such as the volume say), but are {\em insensitive} to the overall radius of the ball. This was implicitly done in \cite{Jacobson2015}, where the variation of area at first order keeping the volume fixed was shown to be equivalent to the variation 
\beq\label{eqn:order1}
\delta^{(1)} A -\frac{d-2}{\ell} \delta^{(1)} V .
\eeq
Here the superindex ${}^{(1)}$ refers to the lowest order variation, linear in the curvature, when the Ricci tensor is not vanishing. One can wonder if the previous variation corresponds to the first-order variation of some geometrical invariant intrinsic to the ball and its boundary. The answer is actually yes, as we can for instance use the following invariant 
\beq\label{eqn:isop}
{\cal I} \equiv A - (d-1) V^{\frac{d-2}{d-1}}\left(\frac{\Omega_{d-2}}{d-1}\right)^{\frac{1}{d-1}}.
\eeq 
Due to the classical isoperimetric inequality \cite{Osserman1978}, ${\cal I}$ is non-negative in $(d-1)$-dimensional Euclidean space, providing a lower bound for the area $A$ of any surface enclosing a given volume $V$ (alternatively, the maximum volume enclosed by a given area). And ${\cal I}$ vanishes only for round spheres enclosing a round ball. Therefore, in our case, at zero order (in flat spacetime) we have for the geodesic ball
\beq
{\cal I}^\flat = 0 
\eeq
and this is independent of the radius of the ball. The first-order variation of ${\cal I}$, linear in the curvature, can be easily computed, leading to
\beq
\delta^{(1)}{\cal I} = \delta^{(1)} A -\frac{d-2}{\ell} \delta^{(1)} V
\eeq
which coincides with (\ref{eqn:order1}) and therefore with the first-order variation of area at fixed volume. We know this is proportional to the energy density of matter at the center of the ball. Hence, $\delta^{(1)}{\cal I}$ vanishes in Ricci-flat spacetimes, and this statement is independent of the ball radius too. In other words, this holds including a non-zero term $\delta\ell_1$ in (\ref{eqn:deltaL}). We can thus consider the second-order variation of ${\cal I}$, including terms quadratic in the curvature. A direct calculation provides
\beq
\delta^{(2)}{\cal I} = \delta^{(2)} A -\frac{d-2}{\ell} \delta^{(2)} V+\frac{d-2}{2\Omega_{d-2} \ell^d} (\delta^{(1)}V)^2 .
\eeq
Bearing in mind that $\delta\ell_1$ is free, from \cite{Jacobson2015} we have $\delta^{(1)}V =\Omega_{d-2}\ell^{d-2} \delta\ell_1$ in Ricci flat spacetimes. The second-order variations of area and volume have been already computed for $\delta\ell_1=0$. For non-zero  $\delta\ell_1$, they receive corrections proportional to $\delta\ell_1^2$ according to
\begin{eqnarray}
\delta^{(2)} V &= &\delta V +\Delta V +\frac{d-2}{2}\Omega_{d-2} \ell^{d-3} \delta\ell_1^2 ,\\
\delta^{(2)} A &=& \delta A +\Delta A+\frac{(d-2)(d-3)}{2}\Omega_{d-2}\ell^{d-4}\delta\ell_1^2
\end{eqnarray}
where $\delta V$, $\Delta V$, $\delta A$ and $\Delta A$ are given in (\ref{eqn:deltaV}), (\ref{eqn:DeltaV}), (\ref{eqn:dA}) and (\ref{eqn:DA}), respectively. Inserting this into the previous formula for $\delta^{(2)}{\cal I}$, all dependencies in $X$ and $\delta\ell_1$ cancel out, leaving an expression which is again insensitive to the overall radius of the ball, given explicitly by
\beq
\delta^{(2)}{\cal I} = \delta A|_V,
\eeq
with $\delta A|_V$ given in (\ref{eqn:dAV}).
In other words, the second order variation of ${\cal I}$ coincides exactly with the second-order variation of area at fixed volume, and they are independent of the radius of the ball. This has a positive side, as one does not have to care about fixing the total size of the ball but, of course, it does not provide the desired answer because, as argued above, such a variation can never be proportional to $W$. We can speculate about the existence of invariants other than ${\cal I}$ which can have the property of independence of the ball radius, and perhaps are also insensitive to the choice of shape deformation, that evaluate to the Bel-Robinson superenergy density $W$ at 
second order in perturbations.

There is a final open possibility that we can contemplate: one could argue that the round ball in flat space is the wrong ball to which to 
compare when considering second order variations.  This  occurs in other quasilocal energy
prescriptions, such as that of  Wang and 
Yau \cite{Wang2009}, which chooses a comparison ball in 
flat space via an isometric embedding.\footnote{Although in our case, the flat space
ball should not be isometric to the curved space one, since then the area variation would vanish.}
Allowing for deformations of the flat space ball yields even more freedom, since now we can 
choose $Y_{ij}$ and $X$ in the curved space independently of their flat-space analogs,  
$Y_{ij}^\flat$ and 
$X^\flat$.  Since only $X-X^\flat$ will be relevant in the comparisons, we now have three free
parameters defining the two balls, which gives enough freedom to fix the volume and find an
area variation proportional to $W$. Actually, in order to avoid the problem with the overall radius we can again use the invariant (\ref{eqn:isop}), which gets rid of the freedom encoded in $X$ and $X^\flat$. By doing so, we can compare the values of ${\cal I}$ in the curved spacetime with respect to a deformed ball in flat space at second order in perturbations. In this fashion, the deficit
\beq
\delta^{(2)}{\cal I} -\delta^{(2)} {\cal I}^\flat 
\eeq
can always be made proportional to $W$ by simply choosing for instance
\beq
(Y^\flat)^2 =\ell^6 E^2\left(\gamma^2 -\frac{2}{3d} \gamma +\frac{5}{18d(d+3)}\right)
\eeq
whose value for $\gamma =1/6$ becomes
\beq
(Y^\flat)^2 =\ell^6 E^2\frac{(d+1)(d-2)}{36d(d+3)} .
\eeq

Unfortunately, what is missing from this procedure is an 
independent justification for choosing the deformations. 
Moreover, the fact that the deformed ball in flat spacetime will no longer be the base of a small causal diamond casts some doubts about the entire argument.

\section{Relation to the Noether charge} \label{sec:noether}
This section seeks to interpret the area of the ball in terms of a Hamiltonian associated with 
evolution within the ball's domain of dependence, 
using the Noether charge formalism \cite{Wald1993, Iyer1994a}.   The ambiguity 
in finding a reference ball to which to compare the area will manifest itself in this 
section in an ambiguity in the zero value of this Hamiltonian.  Furthermore, we will argue that 
the additional pieces beside the Bel-Robinson density $W$ arising in the area can be compensated using ambiguity terms in the Noether charge \cite{Jacobson1994b} depending on the extrinsic curvature of the ball's surface.

\subsection{Second variation of the Noether charge}

We want to demonstrate that the Hamiltonian associated with the spatial ball may be identified
with the Noether charge.  For this we must construct a phase space for describing evolution
within the region.  Starting with a Lagrangian $L[\phi]$, taken as a spacetime $d$-form 
depending on the dynamical fields $\phi$, the equations of motion are obtained by varying the 
Lagrangian with respect to the fields.  We find then that 
\beq \label{eqn:dL}
\delta L = E\cdot \delta \phi + \dd\theta[\delta\phi],
\eeq
where $E=0$ are the dynamical field equations.  The additional term in this variation defines the 
symplectic potential $(d-1)$-form $\theta$.  Taking an antisymmetric variation of $\theta$ defines
the symplectic current,
\beq
\omega[\delta_1\phi,\delta_2\phi] = \delta_2\theta[\delta_1\phi] - \delta_1\theta[\delta_2\phi],
\eeq
and integrating this over the ball gives the symplectic form for the phase space,
\beq
\Omega = \int_\Sigma \omega[\delta_1\phi, \delta_2\phi].
\eeq
Given a vector field $\zeta^\al$, the evolution of the dynamical fields along this vector field is 
generated by a Hamiltonian $H_\zeta$ whose variation satisfies
\beq
\delta H_\zeta = \int_\Sigma\omega[\delta\phi, \lie_\zeta \phi].
\eeq

In a diffeomorphism-invariant theory, this Hamiltonian may be written as a boundary integral
when the equations of motion are satisfied.  Assuming $\zeta^\al$ vanishes on the boundary
$\partial \Sigma$, the variation of the Hamiltonian is simply
\beq \label{eqn:dHz}
\delta H_\zeta = \int_{\partial \Sigma} \delta Q_\zeta,
\eeq
where the Noether charge $Q_\zeta$ is determined by the Lagrangian and symplectic potential
for the theory.  In the case of general relativity, the Lagrangian, symplectic potential, and Noether
charge are given by
\begin{align}
L &= \frac1{16\pi G} R \ep  \\
\theta &= \frac1{32\pi G}(g^{\al\ga} \nabla^{\be}\delta g_{\be\ga} 
- \nabla^{\al}g^{\be\ga}\delta g_{\be\ga} )\ep_\al \label{eqn:theta} \\
Q_\zeta &= -\frac{1}{16\pi G} \nabla^{\al} \zeta^\be \ep_{\al\be},
\end{align}
where $\ep$, $\ep_\al$ and $\ep_{\al\be}$ all denote the spacetime volume form, 
with all but the displayed indices suppressed.  Because $\ze^\al$ vanishes on $\partial \Sigma$,
its covariant derivative there satisfies $\nabla_\al \zeta_\be = g_{\be\ga}\partial_\al\ze^\ga$,  
and  the integral of $Q_\zeta$
over the boundary of the ball can then be expressed as 
\beq \label{eqn:intQz}
\int_{\partial \Sigma} Q_\zeta= -\frac{1}{16\pi G} \int_{\partial \Sigma} \mu \, n\indices{_\ga^\al}
\partial_\al \ze^\ga,
\eeq
where $\mu$ is the induced volume form on $\partial\Sigma$, and $n_{\al\be}$ is the two-form or binormal associated to space normal to $\partial\Sigma$,
defined in terms of a future-pointing, timelike unit normal $u^\alpha$ and an
orthorgonal, outward-pointing spacelike unit normal $N^\beta$ as 
\beq
n_{\al\beta} = 2 u_{[\al}N_{\beta ]}.
\eeq

The geometry under consideration will be taken to be a perturbation of 
a spatial ball in flat space.  The vector $\zeta^\al$ in the flat background 
is chosen to be the conformal Killing vector
generating a flow within the causal development of the ball, 
given in the usual Mikowski coordinates $(t, x^i)$ by
\beq
\zeta^\alpha = \frac{\ell^2 - r^2-t^2}{\ell^2} \partial_t^\alpha -\frac{2t}{\ell^2} x^i\partial_i^\alpha.
\eeq
At the boundary of the ball $r=\ell$, $\zeta^\al$ vanishes, and its derivative satisfies
\beq
\partial_\al \zeta^\ga = \kappa\, n\indices{_\alpha^\ga},
\eeq
where $\kappa = 2/\ell$ is the surface gravity of the conformal Killing vector \cite{Jacobson1993}.  
Using this, we see that the integrated Noether charge (\ref{eqn:intQz}) is simply proportional
to the area 
when evaluated in the background,
\beq
\int_{\partial\Sigma} Q_\zeta = \frac{-\kappa}{8\pi G} A.
\eeq

Next we consider perturbations of the Noether charge.  The perturbations only act on the metric, 
so that $\zeta^\al$ and its partial derivative remain fixed, as does the coordinate position of the 
surface.  The unit normals will also change under the variation in order to remain normalized.  
Their perturbations are thus given by
\begin{align}
\delta u_\al &= -\frac12u_\al u^\be u^\ga \delta g_{\be\ga},  \label{eqn:dua} \\
\delta N_\al &= \frac12 N_\al N^\be N^\ga \delta g_{\be\ga}. \label{eqn:dna}
\end{align}
This also means that the perturbation of the binormal is 
\beq \label{eqn:dnab}
\delta n_{\al\be} = \frac12 n_{\al\be} s^{\mu\nu}\delta g_{\mu\nu},
\eeq
where $s^{\mu\nu}\equiv -u^\mu u^\nu + N^\mu N^\nu$ is the projector orthogonal to $\partial\Sigma$ (the spacetime version of the inverse metric on the normal bundle).  
Additionally, from the fact that the mixed index binormal satisfies $n\indices{_\al^\be} 
n\indices{_\be^\al} = 2$, we see that the perturbations of $n\indices{_\al^\be}$ satisfy
\begin{align}
n\indices{_\al^\be}\delta n\indices{_\be^\al} &= 0 \\
\delta n\indices{_\al^\be} \delta n\indices{_\be^\al}&=-2n\indices{_\al^\be} 
\delta^{(2)} n\indices{_\be^\al},
\end{align}
where $\delta^{(2)} n\indices{_\be^\al}$ 
denotes the change in $n\indices{_\be^\al}$ at second order in 
$\delta g_{\mu\nu}$. 

Applying these identities to the expression for the integrated Noether charge (\ref{eqn:intQz}), we
find that
\beq
\int_{\partial \Sigma} \delta Q_\zeta = -\frac{1}{16\pi G} \int_{\partial \Sigma} \partial_\al\ze^\ga
(n\indices{_\ga^\al}\delta\mu + \mu \delta n\indices{_\ga^\al}) = \frac{-\kappa}{8\pi G} \delta^{(1)} A,
\eeq
so that the Noether charge remains proportional to the area at first order in perturbations. For the vacuum case we are interested in, this variation actually vanishes at this order.

 The 
second order calculation yields
\begin{align}
\int_{\partial \Sigma} \delta^{(2)} Q_\zeta &= -\frac1{16\pi G} \int_{\partial \Sigma} \partial_\al \ze^\ga
(n\indices{_\ga^\al} \delta^{(2)} \mu + \delta n\indices{_\ga^\al}\delta \mu + 
\delta^{(2)} n\indices{_\ga^\al} \mu) \nonumber \\
&= -\frac{\kappa}{8\pi G} \delta A + \frac{\kappa}{32 \pi G} \int_{\partial\Sigma} \mu 
\delta n\indices{_\al^\ga}\delta n\indices{_\ga^\al}, \label{eqn:d2Qz}
\end{align}
where $\delta A$ is the expression given in (\ref{eqn:dA}). Equation (\ref{eqn:d2Qz}) shows that the Noether charge deviates from the area at second order by a term proportional
to the integrated norm of $\delta n\indices{_\al^\ga}$.  We can evaluate this term using 
the form of the binormal perturbation (\ref{eqn:dnab}), which leads to 
\beq
\delta n\indices{_\al^\ga}\delta n\indices{_\ga^\al} = -s^{\al\be}s^{\mu\nu}
\delta g_{\al\mu} \delta g_{\be\nu} +\frac12(s^{\al\be}\delta g_{\al\be})^2.
\eeq
This can be further simplified by noting that the unit normals are orthogonal even after perturbing
the metric, $u_\al N_\be g^{\al\be} = 0$.  Since the perturbations to the unit normals given in 
(\ref{eqn:dua}) and (\ref{eqn:dna}) are each proportional to the original normal, it must be that 
$u_\al N_\be \delta g^{\al\be}=0$, leading to
\beq \label{eqn:dndn}
\delta n\indices{_\al^\ga}\delta n\indices{_\ga^\al} = -\frac12(u^\al u^\be \delta g_{\al\be} + n^\al
n^\be \delta g_{\al\be})^2.
\eeq

Finally, we would like to evaluate this contribution in the gauge set by using Riemann normal 
coordinates, with the first order metric perturbation given by
\beq
\delta g_{\al\be} = -\frac13 x^\ga x^\de R_{\al\ga\be\de}.
\eeq
These coordinates have the useful property that $N^\alpha \delta g_{\al\be}=0$, so that the 
correction to the area term in the Noether charge becomes
\beq
-\frac{\kappa}{64\pi G} \int_{\partial \Sigma} \mu (u^\al u^\be \delta g_{\al\be})^2 = 
-\frac{\kappa}{8\pi G} \frac{\Omega_{d-2} \ell^{d+2}}{d^2-1} \frac{E^2}{36},
\eeq
and the final expression for the second variation of the Noether charge is
\beq \label{eqn:d2Qzgauge}
\int_{\partial \Sigma} \delta^{(2)} Q_\zeta = \frac{\kappa}{8\pi G}\frac{\Omega_{d-2} \ell^{d+2}}{d^2-1} 
\left(\frac{W}{15} -
\frac{E^2}{12}\right).
\eeq

This result shows that the notion of energy provided by the integrated Noether charge differs from
just the area when working to second order in perturbations; however, it still does not produce
a quantity proportional to the Bel-Robinson energy density $W$ at second order in the 
Riemann normal coordinate expansion.  In fact, equation (\ref{eqn:dndn}) shows that the correction
to the area must be nonpositive, which is precisely the opposite of what is needed to cancel the 
unwanted $E^2$ term in the area variation. Note 
also that the correction to the area term provided
by (\ref{eqn:dndn}) depends only on how the metric in the normal directions varies.  In particular, we
could change this term with an active
diffeomorphism $g_{\alpha\beta}\rightarrow g_{\alpha\beta} +\lie_\xi g_{\alpha\beta}$ that fixes the geometry of the surface, and one can
always find such a transformation that causes the metric perturbations in the normal
direction appearing in (\ref{eqn:dndn}) to vanish.\footnote{The reason
that $\delta^{(2)} Q_\zeta$ is not 
completely diffeomorphism-invariant is that it is defined with respect to the fixed vector
$\zeta^\al$, which does not transform under diffeomorphisms.}
In such a gauge, $\zeta^\al$ remains tangent to the null normal surfaces emanating from 
$\partial\Sigma$, unlike in the Riemann normal coordinate gauge.  This is consistent
with the result that the Noether charge will be proportional to the area as long as 
$\zeta^\al$ remains tangent to the null hypersurfaces in the perturbed geometry, which holds at 
all orders in perturbation theory \cite{Hollands2013}.  In particular, if  the geometry of the ball
is such that
$\partial\Sigma$ remains at the intersection of two light cones as described in section
\ref{subsec:diamond}, the Noether charge will simply be given by the area of the ball.  

Finally, note equation (\ref{eqn:dHz}) determines the Hamiltonian to be given by the Noether
charge up to an overall constant, $H_\zeta = \int_{\partial\Sigma} Q_\zeta - H_0$. It is natural
to think of this constant as the area of the ball in flat space, so that the Hamiltonian vanishes
when there are no gravitational fields.  Of course, this necessitates a choice of the flat 
space ball to which we are comparing.  The round sphere in flat space for whom $\zeta^\alpha$
is a conformal Killing vector appears to be a natural choice; however, other choices certainly
are possible.  We thus again face the issue encountered in section \ref{sec:volume} of 
needing to determine an appropriate ball in flat space to compare to, and unfortunately 
our Noether charge arguments do not shed any light on this issue.

\subsection{Extrinsic curvature ambiguities}

There is one loophole that may be exploited to make the integrated Noether charge coincide 
with the Bel-Robinson energy density in the small ball limit.  The Noether charge is subject to 
a number of ambiguities \cite{Jacobson1994b}, one of which arises from an ambiguity
in the symplectic potential $\theta$.  As is apparent from equation (\ref{eqn:dL}), $\theta$ is 
defined only up to the addition of a closed $(d-1)$-form, $\dd\beta[\delta \phi]$.  Adding 
such a term to $\theta$ also changes the Noether charge by
\beq
Q_\zeta\rightarrow Q_\zeta + \beta[\lie_\zeta \phi].  
\eeq
As explained in \cite{Sarkar2013, Wall2015, Bueno2017}, since $\zeta^\al$ acts like a boost
at $\partial \Sigma$ (i.e.\ $\zeta^\al$=0 and $\partial_\al\zeta^\be\propto n\indices{_\al^\be}$),
$\beta[\lie_\zeta\phi]$ will consist of a sum of boost-invariant products of the form $B^{(-m)}\cdot
C^{(m)}$, where $m\neq0$ denotes the boost weight of the tensor. When a tensor is decomposed
into components tangent to the surface and those parallel to the null normals $k_\pm^\al$, 
the boost weight is simply the number of $k_+^\al$ components minus the number of $k_-^\al$ 
components.  

Invariants formed from the extrinsic curvature of $\partial\Sigma$ provide an example of this type
of ambiguity. The extrinsic curvature tensor, usually called shape tensor or second fundamental form vector, can be defined in our case as
\beq
K^\al_{AB}=K_{AB} N^\al - \varkappa_{AB} u^\al\label{eqn:shape}
\eeq
where the second fundamental forms $K_{AB}$ along $N^\al$ and $\varkappa_{AB}$ along $u^\al$ where introduced in sections \ref{subsec:K} and \ref{subsec:nulltheta}, respectively. 
Observe that the $\alpha$ index is purely normal, so 
it will lead to pieces that have boost weights of $\pm 1$.  Hence, the contraction $K\indices{^\gamma
_A_B} K_{\ga CD}$ gives a boost-invariant tensor composed of terms with nonzero boost 
weight, and once the remaining tangential indices are contracted it gives the required form for 
a Noether charge ambiguity. The shape tensor (\ref{eqn:shape}) can be decomposed into its trace, the {\em mean curvature vector}
\beq
K^\al := q^{AB}K^\al_{AB}=K N^\al - \varkappa u^\al \label{eqn:mean}
\eeq
and its traceless part, which is conformally invariant and is called the {\em total shear tensor} \cite{CSV}
$$
\tilde{K}^\al_{AB}:= K^\al_{AB}-\frac{1}{d-2} K^\al q_{AB} , \hspace{1cm} q^{AB} \tilde{K}^\al_{AB} =0 .
$$
Therefore, we can build two independent contractions by
\beq\label{eqn:K2}
K^\gamma K_\gamma, \quad \tilde{K}^\al_{AB} \tilde{K}_\al^{AB},
\eeq
Additional ambiguities involving more factors of extrinsic curvature also arise in this fashion, but
we will restrict attention to the quadratic invariants (\ref{eqn:K2}).  

In principle, this would require to know $K_{AB}$ and $\varkappa_{AB}$ to second order in the curvature. However, we already know that $\varkappa_{AB}$ vanishes in the flat background, which implies that we only need to know it at first order, and this was already given in (\ref{eqn:kappa}). In particular, its trace $\varkappa$ vanishes at this order as shown in (\ref{eqn:kappatrace}).

The situation is different for $K_{AB}$, whose first order expansion is already known (set $Y_{[s]}=0$ for all $s$ in (\ref{eqn:2FF})). But given that its flat space value does not vanish we need to go to next order in the computation. This can be done by noticing ---as in section \ref{subsec:K}--- that $K_{AB}=-\Gamma^r_{AB}$, where $\Gamma$ are the Christoffel symbols of the $t=0$ hypersurface in spherical coordinates, leading after a simple calculation to
\begin{eqnarray}
K_{AB} =\ell  \Omega_{AB}  -\frac{2}{3} \ell^3 n^k n^l D_{ikjl}  \frac{\partial n^i}{\partial\theta^A} \frac{\partial n^j}{\partial\theta^B}+\frac{5}{12}\ell^4 n^k n^l n^m \nabla_k R_{iljm}  \frac{\partial n^i}{\partial\theta^A} \frac{\partial n^j}{\partial\theta^B}  \nonumber\\
+ 3\ell^5 n^k n^l n^m n^p \left(-\frac{2}{45} H_{kil} H_{mjp}+\frac{2}{45} D^q{}_{kil}D_{qmjp} -\frac{1}{20} \nabla_k\nabla_l R_{imjp} \right) \frac{\partial n^i}{\partial\theta^A} \frac{\partial n^j}{\partial\theta^B}. \label{eqn:KAB}
\end{eqnarray}

Concerning the ambiguity term $K^\ga K_\ga$ in (\ref{eqn:K2}), on using (\ref{eqn:mean}) we immediately get
\beq
K^\gamma K_\ga = -\varkappa^2 + K^2
\eeq
and given that $\varkappa$ vanishes at order 1 (and order zero) we only need to compute $K$ up to second order. This can be done from (\ref{eqn:KAB}) and from (\ref{eqn:1FF}) (with all $Y_{[s]}=0$) leading to
\begin{eqnarray}
K = \frac{d-2}{\ell} -\frac{\ell}{3} E_{ij}n^i n^j +\az{\frac{7}{12}} \ell^2 n^k n^l n^m \nabla_kR_{0l0m}\nonumber
\\-\frac{\ell^3}{45}n^i n^j n^k n^l(D\indices{_i^m_j^n}D_{kmln} +4 H\indices{_i^m_j}H_{kml})
\az{-\frac{1}{10} \ell^3 n^k n^l n^n n^n \nabla_k \nabla_l R_{0m0n}} . \label{eqn:K2nd}
\end{eqnarray}

The contribution to the Noether charge will be the integral of this expression over the ball.  
Since the volume element $\sqrt{q}$ 
contains corrections $A_1$ and $A_2$ at first and second
order in the curvature, the full result at second order consists of 
\beq\label{eqn:intK2}
\int_{\partial\Sigma} \mu K_\gamma K^\gamma = \int_{\partial \Sigma} d\Omega_{d-2}\left(2K_2 K_0+
K_1^2+2K_0 K_1 A_1+K_0^2 A_2\right),
\eeq
where $K_{0,1,2}$ denote the value of $K$ at zeroth, first and second order ---the order-$\frac{3}{2}$ term does not contrinute upon integration due to the odd number of $n$'s.
Note that the first order term vanishes since it would involve the trace of $E_{ij}$ after integrating 
over the ball, and the background value is just $\Omega_{d-2} \ell^{d-4} (d-2)^2$.  
Once the integral over the ball is performed, (\ref{eqn:intK2}) evaluates to
\begin{align}
\int_{\partial \Sigma} \mu K_\gamma K^\gamma &= \frac{\Omega_{d-2}\ell^d}{15(d^2-1)} 
\left[-\frac{(d+6)(d-2)}{8}(D^2+4H^2) + \frac{d^2+4d-2}{3} E^2\right] \nonumber\\
&= \frac{\Omega_{d-2} \ell^d}{15(d^2-1)}
\left[ -(d+6)(d-2)W + \frac{5}{6} (d^2+4d-8) E^2 \right] . \label{eqn:K2WE}
\end{align}

This contains the same quadratic Weyl tensor invariants as does the area variation (\ref{eqn:dA}) and the Noether charge
(\ref{eqn:d2Qzgauge}), 
and hence it is possible to choose a combination such that only  a term proportional to 
$W$ survives.  
Such a combination is given by
\beq \label{eqn:areaK2}
H_\zeta = \int_{\partial\Sigma} Q_\zeta + \frac{3 \kappa \ell^2}{16\pi G(d^2+4d-8)}\int_{\partial\Sigma} \mu 
K_\gamma K^\gamma,
\eeq
which at second order in the curvature expansion evaluates to 
\beq
\delta^{(2)} H_\zeta = \frac{-\kappa}{8\pi G} \,\frac{\Omega_{d-2} \ell^{d+2} }{15(d^2-1)} \,  \frac{d^2+4d -20}{2(d^2+4d-8)} W.
\eeq

Finally, concerning the other possible ambiguity term  $\tilde{K}^\al_{AB}\tilde{K}_\alpha^{AB}$, first of all we notice that
\beq
\tilde K^\al_{AB}=\tilde K_{AB} N^\al - \tilde\varkappa_{AB} u^\al\label{eqn:shear}
\eeq
where $\tilde K_{AB}$ and $\tilde\varkappa_{AB}$ are the corresponding shear tensors or trace-free second fundamental forms, so that
\beq
\tilde{K}^\al_{AB}\tilde{K}_\alpha^{AB}=\tilde{K}_{AB}\tilde{K}^{AB}-
\tilde\varkappa_{AB}\tilde\varkappa^{AB}.
\label{eqn:shear2}
\eeq
We already know that $\varkappa_{AB}$ is traceless to this order, and thus $\tilde\varkappa_{AB}=\varkappa_{AB}$ is already given in (\ref{eqn:kappa}). 
On the other hand, from (\ref{eqn:KAB}) one easily sees that $\tilde{K}_{AB}$ vanishes at order zero, meaning that, in flat spacetime, the boundary surface of the geodesic ball is always totally umbilical; for our calculation now this implies that we only need to know $\tilde{K}_{AB}$ up to first order. 
Now, using Eqs.(\ref{eqn:1FF}), (\ref{eqn:2FF}) and (\ref{eqn:K}) we readily obtain 
\beq
\tilde{K}_{AB}=-\frac{\ell^3}{3}  \frac{\partial n^i}{\partial\theta^A} \frac{\partial n^j}{\partial\theta^B}n^k n^l \left(D_{ikjl} +\frac{1}{d-2}(\delta_{ij} -n_i n_j) E_{kl} \right).
\eeq

From this expression together with (\ref{eqn:kappa}) it seems that the ambiguity
term $\tilde{K}^\al_{AB}\tilde{K}_\alpha^{AB}$ will not produce helpful curvature invariants 
when integrated over the ball to give only $W$ when added to the area variation.  This can be deduced from the opposite signs in the squared terms on the righthand side of (\ref{eqn:shear2}), indicating that the coefficients of $H^2$ and $D^2$ will appear in the integrated expression with opposite signs.

\section{Discussion}
This paper has sought to extend the connection between the 
areas of small spheres and energy density
to the case of vacuum general relativity where the matter stress tensor vanishes.  
Since gravity gravitates, we expected a contribution to the area 
proportional to the quasilocal gravitational energy associated with the sphere, the natural 
candidate being the Bel-Robinson super-energy density $W$.  However, as exhibited in equation
(\ref{eqn:dV}), the area variation  
for a geodesic ball depends on other quadratic curvature invariants besides $W$.  
We further 
noted that there is considerable ambiguity in defining the shape of the ball in curved space, and 
each prescription for this shape can lead to a different value for the area variation.  

In section \ref{sec:deformations}, we explored the effect that a general shape deformation has 
on the area, and argued that choices of these deformations exist that cause the area variation to be 
proportional to $W$.  Sections \ref{sec:fixing} and \ref{sec:volume} were devoted to 
exploring various ways of fixing the the shape deformations.  A particularly natural 
choice for the spin-2 deformation was giving by $Y_{ij} = \frac{\ell^3}{6} E_{ij}$, which coincides
with keeping the ball surface at the intersection of two lightcones.  
We also discussed how to vary the radius 
of the ball, and found that the most natural prescription, holding the spatial volume fixed,
precludes an area variation proportional to $W$.  We put forward some ideas on how to 
make progress on the issue of fixing the volume variation in section \ref{sec:volume}. 
 
 We also connected the area variation to the Hamiltonian associated with the ball 
 in section \ref{sec:noether} using the 
 Noether charge formalism.  We showed that there exists a gauge choice for which the 
 Noether charge of the ball coincides with the area, consistent with the results of 
 \cite{Hollands2013}.  This Noether charge has an interpretation as the Hamiltonian generating 
 a flow within the ball's domain of dependence, which  
 justifies associating the area with the quasilocal energy.  Furthermore, there exist ambiguities
 in defining the symplectic form for the ball, which translate to ambiguities in Noether charge and 
 Hamiltonian.  Invariants formed from the extrinsic curvature of the ball are one type of ambiguity,
 and for a specific choice (\ref{eqn:areaK2}) involving $K_\gamma K^\gamma$ one can get the second variation of the 
 Hamiltonian to coincide with $W$.
 Unfortunately, this choice does not appear particularly  natural, since it involves a coefficient that depends explicitly on the background radius of  the ball.  
 
 Despite much effort, we were  unable to make a fully
 satisfactory connection between the area
 and the Bel-Robinson superenergy density.  However, our investigations were by no means 
 exhaustive, and we leave open the possibility that a natural prescription exists for fixing the 
 shape of the ball  that  yields $W$ as the first correction to the area.  We have developed
 a geometric framework for computing area, volume, and extrinsic curvatures perturbatively
 in Riemann normal coordinates with generic shape deformations, which 
 will be of use when investigating other ways of fixing the shape of the ball.  These geometric
 calculations also complement other investigations of small causal diamonds \cite{Gibbons2007, Berthiere2015, Jubb2017}, and may be relevant to ideas in the theory of  causal sets 
 \cite{Bombelli1987}.
 
 There are several future directions to investigate.  The calculation of the shape deformation that 
 ensures that the boundary of the ball lies at the intersection of two light cones was done to first order
 in the curvature at the center of the ball.  One could carry this calculation out to the next order,
 computing the quadratic curvature corrections to the  functions $\phi$ and $\psi$ in 
 (\ref{eqn:phi}) and (\ref{eqn:psi}).  One can argue that these corrections will be a combination of 
 four terms of the form 
 $x^i x^j E\indices{_i^k} E_{jk}$, $x^i x^j x^k E\indices{_i^l} H_{jlk}$, $x^i x^j x^k x^l
 E_{ij} E_{kl}$ and $x^i x^j x^k x^l H\indices{_i^m_j}H_{kml}$.  The first, third and fourth of these 
 affect the area by changing the average value of the radius of the ball, and hence
 have a similar effect as does $X$ from equation (\ref{eqn:DA}).  
 These contributions can be thought 
 of as coming from the focusing of light rays due to the quasilocal gravitational energy within the 
 ball.  Working out these contributions could be a step in the right direction toward getting $W$ in 
 the area variation, although there is still an overall ambiguity in choosing the size of the ball in the 
 curved manifold via the choice of $X$.  
 
 In the first law of causal diamonds, this ambiguity in the ball radius is resolved by fixing the volume
 of the ball.  To arrive at the fixed volume constraint, one compares variations of 
 the off-shell Hamiltonian, given by the symplectic flux through the ball, to its on-shell value, given
 by the area.  It is possible a similar relation fixes the overall size ambiguity at second order 
 in curvature, but this requires evaluating the second variation of the off-shell Hamiltonian, similar
 to the analysis of Hollands and Wald
 \cite{Hollands2013}.  The calculation in the present case is further complicated by the fact 
 that we are perturbing a causal diamond with a conformal Killing vector, while the Hollands and
 Wald calculation involves a true Killing vector.  Another possibly useful construct in fixing the 
 ball shape is to introduce edge mode fields as in \cite{Donnelly2016, Speranza2017}, since these 
 tend to parameterize all possible ways of deforming the boundary of a local subregion.  It is 
 possible that these could lead to a combination of geometric invariants that are independent of the 
 way the shape of the ball is chosen.  This would be similar to how the ``isoperimetric'' invariant (\ref{eqn:isop}) is independent, at zero, first, and second order at least, of changes in the overall radius, as argued in section \ref{sec:volume}.
 
Since the conformal Killing vector  vanishes at the boundary of the ball, the 
Hamiltonian we are computing corresponds to a flow that remains within the domain of dependence
of the ball.  It may be that the  quasilocal energy corresponding to $W$ should be 
conjugate to a flow that is not vanishing at the boundary of the ball, such as the 
usual time translation
vector.    
In the Noether charge 
formalism, this produces an additional boundary term in the Hamiltonian depending on 
$\xi\cdot\theta$, where $\xi$ is the nonvanishing vector field for the flow, and $\theta$ is the 
symplectic potential (\ref{eqn:theta}).  However, identifying this with a Hamiltonian is subtle because
$\xi\cdot\theta$ is in general not integrable, i.e.\ it cannot be written as $\delta \mathcal{B}$ for some 
covariant $(d-2)$-form  $\mathcal{B}$.  The issue is that symplectic flux can leak out through the region
when evolving along the flow if $\zeta$ does not vanish at the boundary \cite{Wald2000a}, unless
boundary conditions are imposed on the dynamical fields.  
Thus, it remains to be seen whether this gives a useful notion of energy in the small ball
limit.

 Finally, we note that all of these calculations were performed for arbitrary 
 spacetime dimension.  Much of the work on quasilocal mass has focused on $d=4$, where the 
 Weyl tensor is simpler than in higher dimension.  This is because the traceless 
 magnetic-magnetic part of the Weyl tensor, $F_{ijkl}$, vanishes identically in 4 dimensions.  
 Thus, there are only two independent quadratic invariants formed from the electric and magnetic
 parts that can appear in the calculations, and so it is easier to find a prescription for the quasilocal
 energy and the shape that produces the Bel-Robinson superenergy density.  Using the
 machinery developed in this paper, one could investigate various prescriptions for the 
 quasilocal mass in higher dimensions, where their ability to reproduce $W$ in the small 
 ball limit would be more nontrivial.  It is worth noting that some prescriptions for quasilocal
 mass are not even extendible to higher dimension, e.g.\ the isometric embeddings into flat space
 used for the Wang-Yau quasilocal mass \cite{Wang2009} 
 will generically not be possible for codimension-2 
 surfaces in higher dimension.  Since gravity in higher dimensions has many interesting 
 applications, a dimension-independent notion of quasilocal gravitational energy is worth pursuing.

\section*{Acknowledgments}
J.M.M.S. is supported under Grants No. FIS2014-57956-P (Spanish MINECO-fondos FEDER), No. IT956-16 (Basque Government), and EU COST action No. CA15117 ``CANTATA.''  T.J. and A.J.S. 
are supported by the National Science Foundation under grants No.\ PHY-1407744 and PHY-1708139.  A.J.S. acknowledges support from the Monroe H. Martin Graduate Research 
Fellowship.

\appendix

\section{Corrections in v2} \label{sec:corr}

A computational mistake occured  in  v1 of this paper. Though the main conclusions and results do not change, there are some sentences and explanations that have been corrected in the 
present version.   The main error occurred in equation (\ref{eqn:Gamma0}), which in v1 did
not contain the final term involving $\tilde\delta\ell_1$ 
in the expression for $\stackrel{1}{\Gamma^r_{AB}}$. 
This error then affects equations (\ref{eqn:2FF}-\ref{eqn:YK}), which have been corrected in
v2.  The discussion in the paper has been updated to reflect these changes, and we note that 
the primary difference in the discussion is that holding the trace of the
 extrinsic curvature $K$ fixed 
generically produces a different shape variation of the ball than demanding that the 
ball arise as the intersection of two lightcones, whereas in v1 we claimed that these two 
procedures gave the same shape deformation to first order in RNC.

\bibliographystyle{JHEP}
\bibliography{volume_var}

\end{document}